\documentstyle[12pt]{article}

         \def\be{\begin{equation}}
         \def\bea{\begin{eqnarray}}
         
         \def\o{\over}

         \def\ee{\end{equation}}
         \def\eea{\end{eqnarray}}
         \def\R{\rm {I\kern-.200em R}}
         \def\C{\rm {I\kern-.520em C}}
         \def\t{\tilde}
\begin{document}
\begin{titlepage}
\vspace*{5mm}
\begin{center} {\Large \bf Coulomb gas representation of quantum Hall
 effect\\
\vskip 0.35cm
 on Riemann surfaces}\\
\vskip 1cm
M. Alimohammadi \footnote {e-mail:alimohmd@netware2.ipm.ac.ir} and
H. Mohseni Sadjadi \footnote {e-mail:sadjadi@netware2.ipm.ac.ir}\\
\vskip 1cm
{\it  Physics Department, University of Teheran, North Karegar,} \\
{\it Tehran, Iran }\\

\end{center}
\vskip 2cm
\begin{abstract}

Using the correlation function of chiral vertex operators of the
Coulomb gas model, we find the Laughlin wavefunctions of quantum
Hall effect, with filling factor $\nu =1/m$, on Riemann sufaces
with Poincare metric. The same is done for quasihole wavefunctions.
We also discuss their plasma analogy.

\end{abstract}
\end{titlepage}
\newpage

\section{ Introduction }
Study the behaviour of electrons living on a two dimensional surface and
interacting with a constant magnetic field orthogonal to the surface, is
one of the most important part of physics, known as the quantum Hall effect (QHE).
When the surface is plane, both integer QHE and fractional QHE have been
observed experimentally, and have been described by Landau and Jastrow-type
wavefunctions, respectively [1]. Also by introducing new types of many body
condensates which carry fractional charge, ${\it i.e.}$ anyons, an unified picture
of integer and fractional QHE has been given. Furthermore, as it has been
stressed by
Laughlin, the Laughlin-Jastrow wavefunctions, both for electrons and anyons
have a natural interpretation  in terms of a two dimensional plasma of charges,
interacting by Coulomb forces and embedded in a uniform neutralizing background
[1].

A particular intriguing and interesting case occurs when the two dimensional
surface is a Riemann surface of higher genus. Although it is not accessible
experimentally, the problem of the physics on Riemann surfaces has deep relation
with some interesting problems, like the occurance of chaos in the surface
with negative curvature [2], and recent developments in the theory of surfaces,
for example, the moduli of surface and the vector bundle defined on the moduli [3].
Until now, the QHE on different non-flat surfaces have been studied. For example
on the sphere [4], torus [5], and on the hyperbolic plane [6,7]. A recent and detailed
investigation has been done in ref.[8] which the Landau and Laughlin levels are
studied on Riemann surface with some particular metrics.

In [8], the authors have shown that the wavefunctions consist of two parts.
A holomorphic part which is independent of metric and a known metric--dependent
function. For Landau levels, they have shown that the holomorphic part
is the Slater determinant of sections of the holomorphic line bundle, and for
the Laughlin states they have an ansatz for the holomorphic part and it
has been shown that this function is the determinant of the holomorphic sections
of a vector bundle. This vector bundle is the tensor product of a line bundle
and a flat vector bundle of rank $m$ ($\nu = 1/m$ is the filling factor).

Now, there is another approach for studying the QHE in which the conformal
symmetry of QHE is used to calculate the different quantities.
There are several evidences for the existance of this symmetry.
For example in ref.[9],
it has been shown that the Laughlin wavefunctions are related to the
conformal blocks of two dimensional conformal field theories (CFT).
In the same paper, the fractional statistics of quasiparticles has been
related to braiding properties of the vertex
operators of the Coulomb gas model, which has conformal symmetry.
In ref.[10], it has been shown that the Halperin-Haldane singlet quantum Hall
effect wavefunction can be split
into two parts. One part is related to a state describing a one
 component plasma (OCP) system, and the other part behaves like a
conformal block of primary fields of the $su(2)$ Wess-Zumino-Witten model.
Also in ref.[11], it has been shown that the holomorphic part of the Laughlin
wavefunction on the torus, can be obtained by the correlation function of
the Coulomb gas vertices. Moreover, it has been pointed out that the Coulomb gas
approach and the OCP description of the Laughlin wavefunction are
consistent.

In this paper we will study the QHE  on arbitrary Riemann surface, in the
context of CFT. Our purposes for this investigation are as follows.
First to see that can this relation between QHE and Coulomb gas model be generalized
to general Riemann surfaces? Second, as we will see, this approach is much
easier than those which has been considered in [8]. And third, our approach
can be easily  generalized to the case where anyons also exist.

The plan of this paper is as follows.
In section 2 we will discuss in brief the relation between OCP and CFT
description of QHE. In section 3, by using the correlation function of the
Coulomb gas vertices on Riemann surface (derived in [12-15]), we obtain the
holomorphic part of the Laughlin wavefunction. To do this, we must
appropriately determine the parameters of the corresponding Coulomb
gas model. We will also discuss the  different aspects of this equivalence.
In section 4, we obtain an expression for quasiholes wavefunctions and
determine their charges in this context.

\section{ QHE on the plane}
To obtain an insight about the relation between QHE and CFT, let us recall
the Laughlin wavefunction. It has been shown by Laughlin [1], that the
wavefunction of the QHE is the many particle wavefunction which looks like:
\be
\psi (z_1,{\bar z}_1,\cdots ,z_N,{\bar z}_N )=\prod_{i=1}^N
e^{-\int A_{{\bar z}_i}d{\bar z}_i}F(z_1,\cdots ,z_N).
\ee
$z_j=x_j +iy_j$, is the position of $j$-th particle and
$A_{\bar z}= {1\over 2}(A_x + iA_y)$ is the gauge
potential. The main purpose of theoretical investigations of QHE is to
determine the holomorphic function $F(z_1, \cdots ,z_N)$ which must obey the Fermi
statistics. Laughlin chose this function to be the eigenfunction of angular
momentum, and showed that $\prod_{i<j}^N(z_i-z_j)^m$ is appropriate function for
$\nu=1/m$ filling factor. The final result, in $A_x= -{B\over 2}y$ and
$A_y={B\over 2}x$ gauge, is
\be
\psi (z_1,{\bar z}_1,\cdots ,z_N,{\bar z}_N )=\prod_{i<j}^N(z_i-z_j)^m
\prod_{i=1}^Ne^{-{1\over 4}|z_i|^2},
\ee
By introducing the classical potential energy $U$ through
$ |\psi |^2=e^{-\beta U}$, where
$ \beta^{-1} $ is an arbitrary effective temperature,
Laughlin showed that this system is equivalent to a two dimensional plasma
of particles with electric charge $m$, interacting by Coulomb forces and
embedded in a uniform neutralizing background.

Now the interesting point is that, this OCP description of QHE can also be
achieved by considering the Coulomb gas model.
This model is a free massless
scalar field which modified with
a background charge at infinity. The two point function of
these fields satisfies in [16]
\be
\partial_z \partial_{\bar z} < \Phi (z,{\bar z})\Phi (w,{\bar w})>=\pi
\delta^2(z,w),
\ee
and $\Phi (z,\bar{z})$ splits into holomorphic and antiholomorphic parts. Note
that this equation is nothing but the Laplace equation for the Coulomb interaction.
If $\varphi (z)$ is the holomorphic part of $\Phi (z,{\bar z})$, then the
expectation value of the product of vertex operators
$:e^{iq\varphi (z_i)}:$ is
\be
F(z_1,\cdots , z_N)=<:e^{iq_1\varphi (z_1)}:\cdots :
e^{iq_N\varphi(z_N)}:>=e^{-q_iq_j \sum_{i<j}^N<\varphi (z_i)\varphi
(z_j)>}.
\ee
Now as on the plane $<\varphi(z_i) \varphi(z_j) >= -{\rm ln} (z_i -z_j)$, which is the
Coulomb potential on two dimension, the holomorphic
part of eq.(2) is recovered by choosing $q_i= \sqrt m$ for all $i's$ (as the
particles are identical) and $\beta= 1/m$.
In this manner the Coulomb gas and QHE relate to each other on the plane,
that is each vertex corresponds to an electron and its conformal charge
$\sqrt m$
relates to the electric charge of particles of plasma. {\it In summary} as the
Coulomb gas model is effectively a theory of Coulomb interaction and QHE
has a plasma analogy which again is based on Coulomb interaction, therefore
the result of these two theories {\it coincide}. In the next section we will use
this correspondence and also the braiding properties of vertex
operators, to find the Laughlin wavefunctions on Riemann surfaces.

\section{ Coulomb gas approach of QHE on Riemann surface with Poincare
metric}
Now we study the QHE on a two dimensional compact and orientable Riemann
surface $\Sigma$. On this surface, the charged particles interact with constant
orthogonal magnetic field produced by monopoles. We choose, as in [8], the
Poincare metric $g_{z {\bar z}}= y^{-2}$. The simply connected covering space
of $\Sigma $
is the upper half plane $H$ and $\Sigma= H/ \Gamma $, where $\Gamma$ is a
discrete subgroup of the isometry group of $H$. $\Gamma$ is generated
by Fuchsian transformations around canonical homology basis.
For a covariantly constant magnetic field $B$, and in the symmetric gauge
$A_z= A_{\bar z} = {B\over 2}y$, the one particle Hamiltonian is [8]
\be
H= -g^{z{\bar z}}D{\bar D} +B/4,
\ee
where $D=\partial -{B \over 2} \partial {\rm ln}g_{z{\bar z}}$ and ${\bar D}
= {\bar \partial }+ {B \over 2} {\bar \partial} {\rm ln}g_{z{\bar z}}$ and we
take the electron mass $m=2$ for simplicity.
The ground state wavefunction satisfies in
\be
{\bar D} \psi=0,
\ee
with solution $\psi (z, {\bar z})= {g_{z {\bar z}}}^{-{B\over 2}} F(z)=
y^{B} F(z)$, where $F(z)$ is a holomorphic function. The behaviour of $F(z)$ under Fuchsian transformation have been
discussed in [8]. But here, we want to solve this problem in the context
of CFT, so we need to find the behaviour of the wavefunction under a larger
transformation, that is the general conformal transformation which the
Fuchsian transformation is a subclass of it.
To do so, we note that for two dimensional surface with Poincare metric,
the conformal transformation, which leaves the metric invariant up to
a scale change
\be
g_{{\tilde z}{\tilde {\bar z}}} d{\tilde z} d{\tilde {\bar z}}=
\Omega g_{z{\bar z}} dz d{\bar z},
\ee
reduces to analytic coordinate transformations
\be
{\tilde z}= f(z) \ \ \ \ ; \ \ \ \ {\tilde {\bar z}}= {\bar f}({\bar z}),
\ee
where $f({\bar f})$ is a holomorphic ( antiholomorphic) function.
Under conformal transformation $D$ and ${\bar D}$ change as
\be
{\t D}={dz \o {d{\t z}}}U^{-1}DU \ \ \ \ ;  \ \ \ \
{\t {\bar D}}={{d{\bar z}} \o {d{\t {\bar z}}}}U'^{-1}{\bar D}U',
\ee
where
\be
U(z,{\bar z})=\Omega^{-B/2}\left( {dz \o {d{\t z}}} \right)^{-B/2}
\left( {{d{\bar z}} \o {d{\t {\bar z}}}} \right)^{B/2} \ \ \ \  ; \ \ \ \
U'(z,{\bar z})=\Omega^{B/2}\left( {dz \o {d{\t z}}} \right)^{-B/2}
\left( {{d{\bar z}} \o {d{\t {\bar z}}}} \right)^{B/2}.
\ee
The Hamiltonian (5) in new coordinate is
\be
H=-g^{{\t z}{\t{\bar z}}}{\t D}{\t {\bar D}}+B/4,
\ee
and the transformed ground state wavefunction ${\tilde \psi }$ satisfies in
${\tilde {\bar D}}{\tilde \psi}=0$. Using (6) we find ${\psi = U'{\tilde \psi}}$
and then using (10) we obtain:
\be
{\t \psi }\Omega^{B/2}d{\t z}^{B/2}d{\t {\bar z}}^{-B/2}=
\psi dz^{B/2}d{\bar z}^{-B/2}.
\ee
Now considering the decomposition ${\psi (z,{\bar z})}= y^{B} F(z)$, and putting it
in (12), we find that $F(z)$ must be a primary field of weight $B$, ${\it i.e.}$
a $B$ -form under general conformal transformation.

As was mentioned in introduction, the authors of [8] have found the holomorphic
part of Landau and Laughlin wavefunction by lengthy calculations.
Here we want to calculate these functions by using the plasma analogy of QHE,
that is using again the Green functions of Coulomb gas, but now on Riemann
surface. So let us first bring a quick review about the Coulomb gas model on
Riemann surface. This model is defined by
a bosonic scalar field coupled to a background charge $Q$ and described
by the following action [15]
\be
S={1\over 2\pi}\int d^2z(\partial \Phi (z,{\bar z}) {\bar \partial}\Phi
(z,{\bar z}) +{1\over 4}Q\sqrt{g}R\Phi (z,{\bar z})),
\ee
where $R$ is the scalar curvature of the surface and $g={\rm det}
g_{\mu\nu} $.
In the following we shall only consider the holomorphic part of the
correlation functions and hence we require $\varphi$ ( the holomorphic
part of $\Phi$) to compactify on
a unit circle $R/ 2\pi Z$ [15]. $R$ is real line and $Z$ denotes integer
numbers. The correlation function of vertex fields
$ < \prod_{j=1}^N :e^{i \alpha_j \varphi(z_j)}:>$ is calculated in
different context [12-15]. In [12], it is obtained
by successive application of Wick theorem and by considering the effect
of zero modes.
In [13], it is shown that this correlation function can be derived by
splitting ${\varphi(z)}$ to its zero and nonzero mode components
\be \varphi (z)=2\pi\sum_{i=1}^gp_i\int\omega_i(\nu)d\nu+{\hat
\varphi}(z), \ee
where $p_i$ and ${\hat \varphi}(z)$ are independent free fields and
$p_i$ are
zero mode oscillators. The contraction rule for ${\hat \varphi}(z)$
is $<{\hat
\varphi }(z) {\hat \varphi }(w)>=-{\rm ln} E(z,w)$ [13,15], which is
the Green function
of two charges located at $z$ and $w$, interacting via a
Coulomb potential in two dimension. In [14], the correlation function
is obtained
by using the $b-c$ system, which is described by the first order action
$ S=\int d^2z
 b {\bar \partial} c$. $b$ and $c$ are conformal fields with weights $ \lambda$
and $\lambda -1$,
respectively. By calculating the correlation function of the vertex
fields (vertex insertions), the authors of [14]
have shown that they are the same as the corresponding
one in the Coulomb gas model. The result which is obtained in all above
papers is
\be < \prod_{j=1}^NV_{q_j}(z_j)>=< \prod_{j=1}^N:e^{iq_j\varphi(z_j)}:>=
 \prod_{k=1}^N\sigma^{Qq_k}(z_k)\prod_{i<j}^NE^{q_iq_j}(z_i,z_j)\theta
\left[ \begin{array}  {c} \delta \\ \epsilon \end{array} \right] (cv|d
\Omega).\ee
${\sigma (z)}$ is a holomorphic $(g/2,0)$-form, without zero or pole,
where $g$
is the genus of the surface and ${\sigma (g=1)}=1$. $E(z_i,z_j)$ is a
holomorphic $(-1/2,0)$-form, which is antisymmetric under interchanging
of its coordinates and is zero for
${z_i = {\gamma (z_j)}}$;$\gamma \in \Gamma $, $\Gamma \subset $ PSL(2R).
$v={\sum_iq_iz_i -Q \Delta}$, in which the Riemann class $\Delta $ is a
$(g-1)$
degree divisor. Theta characteristics $(\delta ,\epsilon )$ and $c$
and $d$ must
be consistent with the boundary condition of $V_{q_i}(z_i)$. By boundary
condition we mean the behaviour of $<V_{q_i}(z_i)>$ under winding the point
$z_i$ around the homology cycles of our Riemann surface.
It can also be shown
that the correlation function (15) vanishes, unless the total
charges $q_i$
cancel the background charge $Q$ [12]
\be \sum_iq_i=-{Q \over 8\pi}\int d^2z\sqrt{g}R(z)=Q(g-1).\ee
Now if we want the correlation function (15) describes a fermionic
wavefunction,
$q_i q_j$ must be an odd integer (as $E(z_i,z_j)$ is antisymmetric).
Also if we demand
that all fermions are identical, we must choose all $q_i$'s equal to
$ {\sqrt m} $, where $m$ is an odd integer. In this way
$<\prod_{i=1}^N  V_{q_i}(z_i)> $
becomes a Jastrow type wavefunction.

Another necessary condition for $< \prod_{i=1}^N V_{q_i}(z_i)>$ to be a
Laughlin
wavefunction, is that its behaviour under the action of
conformal
group transformation must be consistent with the conformal
weight of electrons wavefunctions, which as was mentioned after eq.(12)
is equal to $B$.
Now as the conformal weight of $e^{iq\varphi }$ is ${1\over 2}q(q+Q)= {1\over 2}\sqrt{m}
(\sqrt{m} +Q)$, we obtain
\be B={(m+\sqrt{m} Q)\over 2}. \ee
Following the above
discussion, the appropriate wavefunction for a Laughlin state, which satisfies
in eqs.(12) and (16) for each of its coordinates, is
$$ \psi (z_1,\cdots , z_N)= \prod_{i=1}^Ny_i^B< \prod_{j=1}^NV_{q_j}
(z_j)>$$
\be = \prod_{i=1}^Ny_i^B \prod_{i=1}^N \sigma^{2B-m}
(z_i)\prod_{i<j}^N E^m(z_i,z_j)
\theta \left[ \begin{array} {c}\delta \\ \epsilon \end{array} \right]
 (m \sum_{i=1}^N z_i -(2B-m) \Delta |m \Omega ). \ee
Following the freedom of choosing the characteristics ofthe theta
function of eq.(15)
(as discussed in [14]), we can choose $c=\sqrt m$ and $d=m$ in our case.
This choice of $c$ and $d$ is consistent with the wavefunction on the torus [5,11],
and also ensures that the phase of $\psi$, when the points $z_i$'s
wind around the homology cycles, does not depend on $z_i$ . This independence
comes from the
invariance of eq.(6) under this windings [8]. Now what are the characteristics
$\delta$ and $\epsilon$ ? As was discussed in [8], by comparing the behaviour
of wavefunction under Fucshian transformations $z\longrightarrow \gamma z$,
with behaviour of theta functions under similar transformations, one leads to
$\delta= \delta_0 +l/m,(l_i=1,\cdots ,m$ and $i=1,\cdots ,g)$ and $\epsilon= \epsilon_0$,
where $\delta_0$ and $\epsilon_0$ are $g$-component constant vectors with
components in the interval $[0,1)$. These values of $\delta$ and $\epsilon$
give the correct
degeneracy number of the Laughlin wavefunctions, that is $mg-g+1$. The explicit
values of $\delta_0$ and $\epsilon_0$ depend on our explicit choice of the phase
which appears from wavefunction under $z\longrightarrow \gamma z$. For example
in [5,24], these values have been fixed by choosing $u(\gamma_j, z_j)=e^{i\phi_j}
(j=1,\cdots,2g)$, where $u(\gamma ,z)$ is defined through $\psi (\gamma z)=
u(\gamma ,z)\psi (z)$, and $\gamma_j$ is a transformation identifying sides
in the fundamental polygon which represents our Riemann surface in covering space,
and $\phi_j$ is the flux through the $j$-th cycle.
In this way our final result (18) becomes exactly the same as one has been
obtained in [8].

Now let us investigate the plasma description of the the wavefunction (18).
We write
this wavefunction as $\psi = \psi_1 \psi_2 $ where
\be \psi_1 =\prod_{i=1}^Ny_i^B
\prod_{i<j}^NE^m(z_i,z_j),\ee
and
\be \psi_2 = \prod_{i=1}^N \sigma^{2B-m} (z_i)\theta
\left[ \begin{array} {c}\delta \\ \epsilon \end{array} \right] (m\sum_{i=1}^N
z_i-(2B-m)\Delta |m\Omega).\ee
$\psi_1$ only depends on interaction part of the wavefunction,
${\it i.e.}$ Coulomb interaction, and $\psi_2$ is related to spin structure
of the electrons wavefunction on this surface.
Therefore the interaction potential $U$, which can be defined in
\be |\psi_1 |^2= e^{-\beta U},\ee
becomes
\be U=-{1\over \beta}(\sum_{i=1}^N {\rm ln}{y_i}^{2B}+
 \sum_{i<j}^N {\rm ln}{|E(z_i-z_j)|^{2m}}) .\ee
Now as in the fundamental domain of the Riemann surface we have
\be \partial_z \partial_{\bar z}{\rm ln}|E(z,w)|^{2m}=\pi m
\delta^2(z-w),\ee
by choosing ${1 \over \beta }= m$, we see that $U$ is a Coulomb
potential of particles with charge $m$, interacting with themselves and
 with a uniform background charge $\rho_0 = B / 2 \pi $ [7]. Charge
neutrality of the plasma requires that the plasma particles
spread out in the surface with density $\rho_m = \rho_0 / m$, which
corresponds to filling factor $\nu  = 1/m $.

To determine the precise value of $m$, we use eqs.(16) and (17) to obtain:
\be m(N+g-1)=2B(g-1). \ee
This equation gives the value of $m$ in terms of magnetic field $B$, the genus
$g$, and the electrons number $N$. It is also interesting to see the geoemetrical
meaning of eq.(24).
By Riemann vanishing theorem, the number of zeros of the theta function
of eq. (18) is
$mg$ and as $\prod_{i<j}^NE(z_i,z_j)^m$ (as a function of $z_i$), has $m(N-1)$
zeros, so the number of
zeros of the wavefunction (18), with respect to each of its coordinates,
is $mg+m(N-1)$, which from eq.(24) is equal to the magnetic flux $\phi
=2B(g-1)$.
This shows that the degree of line (for $m=1$) or vector
(for $m>1$)
bundle is equal to the first Chern number of the gauge field, as we have expected.

As a last point, we know that the Coulomb gas model, by suitable
choosing of $Q$, can be considered as minimal models.
The minimal models characterized by two positive coprime integers
$p$ and $q$ with central charge
$c(Q)= 1-6(p-q)^2/pq$. Now as the central charge of Coulomb gas
is $ c(Q)= 1-3Q^2$,  eq. (17) shows that if $B$ and $m$ satisfy in
\be ({{2B-m}\over \sqrt m})^2= {2(p-q)^2\over {pq}}, \ee
our QHE is a $(p,q)$ minimal model. For example for $p=q+1$ unitary
minimal models, any odd integer $m$ which satisfies in
\be m= {q(q+1) \over 2} ({r\over s})^2, \ee
where $r$ and $s$ are integers, has a $(q+1,q)$ corresponding
minimal model description. At $Q=0$, $c$ is equal to one and $B$
is $m/2$. A detailed discussion about this case can be found in [18].

\section{ Quasiholes on Riemann surface }

In this section we want to study the aspect of quasihole states in
the context
of conformal field theory. Laughlin argued that the lowing excited
state of QHE
are produced by creation of quasiparticles (quasiholes) in the system.
These are
 fractional statistics particles, ${\it i.e.}$, by interchanging two of them, the
wavefunction takes the $e^{i\theta }$ phase.
This phase for quasiholes is ${\pi /m }$ where
$\nu ={1/m}$ is the filling factor [1,17]. If we want to express the
fractional statistics particles in terms of vertex fields, we must
choose the
appropriate charges for these vertices . Using (15), it is seen that by
interchanging two vertices we obtain
\be <V_{q_i} (z_i) V_{q_j} (z_j)>= e^{i\pi q_i q_j}<V_{q_i} (z_j)
V_{q_j} (z_i)>, \ee
so we must choose $q_i=1/{\sqrt m} $ to relate the vertex fields to the quasiholes.
Now consider a system containing $N$ electrons
( represented by vertices with charge ${\sqrt m}$ ), and $N_q$
quasiholes
( represented by vertices with charges $1/{\sqrt m}$ ), then eq. (16)
leads to
\be N\sqrt m + {N_q\over {\sqrt m }}= Q(g-1). \ee
Using eqs. (16) and (17) ( which also holds in this case), we determine
the filling factor
\be m(N+g-1) + N_q= 2B(g-1)=\phi .\ee
This relation is consistent with the result pointed out in [17],
and can be used to obtain the electric charge of quasiholes
with the method
that was introduced in [4]. If the system of $N$ electrons, in
Laughlin state, is excited by removal
of an electron, at fixed magnetic field, the final state has
the following flux
\be \phi (N;m)= \phi (N-1;m)+m, \ee
where $\phi (N;m)=m(N+g-1)$. Comparing eqs. (29) and (30), showing
that the
new system (30) is composed of $N-1$ electrons and $m$ quasiholes.
Hence the quasiholes
carry the charge $e^*=e/m$ ($e>0$).
To reproduce this result in another way, we note that the charge of
particles
can also be determined by using the OPE of the current and corresponding
fields [9]. On Riemann surface, the above OPE is [19]
\be J(z)e^{i \varphi (w)/{\sqrt m}}= {{1/m}\over z-w}e^{i\varphi (w)
/{\sqrt m}}+\cdots .\ee
Therefore following the steps of [9], the charge of quasihole
corresponding to
vertex $e^{i\varphi (w)/ {\sqrt m}}$ is $e*=e/m$.

At the end, we present an expression for the holomorphic part of the
wavefunction containing $N$ electrons and one quasihole
$$  \psi (z,z_1,\cdots , z_N)= < V_q(z)\prod_{i=1}^N V_{q_i}(z_i)>=
\sigma^{(2B-m)/m}(z)
\prod_{i=1}^N \sigma^{2B-m}(z_i)$$
\be \times \prod_{i=1}^N E(z_i,z) \prod_{i<j}^N{E^m(z_i,z_j)}
 \theta \left[ \begin{array}  {c} \delta \\ \epsilon \end{array} \right]
(m{\prod_{i=1}^N z_i}+z-(2B-m)\Delta |m \Omega),\ee
which is obtained by eq. (15).
Factorize this wavefunction as $\psi =\psi_1 \psi_2$, where
\be \psi_1 =\prod_{i=1}^N E(z_i,z_j)
\prod_{i<j}^NE^m(z_i,z_j),\ee
and $\psi_2 $ other terms,
and again by considering $|\psi_1|^2= {\rm exp }(-U/m)$, one can see
that $U$ is
the Coulomb potential of a system of particles of charge $m$,
interacting with
themselves and with a particle of charge one located at $z$
( which again proves that $e^*=e/m$ ) .

At the end we would like to add a point.
One of the important point in the physics of QHE, is to understand
the incompressibility feature of the Laughlin wavefunctions, which
may be related to quantum group symmetry of the Laughlin states
[7,20-22]. On the other hand, there is a deep connection between
the conformal and quantum group symmetries [23]. Our procedure in
expressing the Laughlin states in the context of CFT, may shed some
light on these connections on Riemann surfaces. We will discuss these,
elsewhere.

\section{ Conclusion }

Using the analogy of the Coulomb gas and plasma description of QHE,
the conformal symmetry of Laughlin states, and the results that have been
found for the Coulomb gas model on Riemann surface, we obtain the Laughlin
wavefunction on an arbitrary compact and orientable Riemann surface.
We have also determined the filling factor and degeneracy of these
wavefunctions. In the case of Poincare metric, we find the plasma description
of QHE on these surfaces and also state the relation between FQHE and
minimal models. Finally, for the cases where the quasiholes are also present,
we find the wavefunctions.
\vskip 1cm
\noindent{\bf Acknowledgement}

We would like to thank the Institute for Studies in Theoretical Physics and
Mathematics, for partial financial support.


\vskip 1cm

\begin{thebibliography}{99}
\bibitem [1]{a} R. Prange, S. Girvin {\it The Quantum Hall Effect}.
2nd edn ( Springer New York, Springer, 1990 ).
\bibitem [2]{b} M. C. Gutzwiller, {\it Chaos in Classical and Quantum Mechanics}
( Springer--Verlag, 1990).
\bibitem [3]{c} J. Fay, American Mathematical Society, Memories, No. {\bf 464}
( Providence, Rhode Island, 1992 ).
\bibitem [4]{d} F. D. M. Haldane; Phys. Rev. Lett. {\bf 51} (1983) 605.
\bibitem [5]{e} F. D. M. Haldane, E. H. Rezayi; Phys. Rev. Lett. {\bf B31} (1985) 2529.
\bibitem [6]{f} A. Comtet; Ann. Phys. {\bf 173} (1987) 185.
\bibitem [7]{g} M. Alimohammadi and H. Mohseni Sadjadi; J. phys. A.
{\bf A29} (1996) 5551.
\bibitem [8]{h} R. Iengo and D. Li; Nucl. Phys. {\bf B413} (1994) 735.
\bibitem [9]{i} G. Moore and N. Read ; Nucl. Phys. {\bf B360} (1991) 362.
\bibitem [10]{j} A. Balatsky and E. Fradkin ; Phys. Rev. {\bf B43} (1990) 10622.
\bibitem [11]{k} G. Cristofano, G Maiella, R. Musto, and F.Nicodemi; Phys. Lett.
{\bf B262} (1991) 88.
\bibitem [12]{l} E. Verlinde and H. Verlinde; Phys. Lett. {\bf B192} (1987) 95,
 Nucl. Phys {\bf B288} (1987) 357.
\bibitem [13]{m} T. Eguchi and H. Ooguri; Phys. Lett. {\bf B187} (1987) 127.
\bibitem [14]{n} L. Bonora, M. Matone, F. Toppan, and K. Wu; Nucl. Phys.
{\bf B334} (1990) 717.
\bibitem [15]{o} O. Lechtenfeld; Phys. Lett. {\bf B232} (1989) 193.
\bibitem [16]{p} P. Christ and M. Henkel; Lecture Notes in Physics {\bf m16} (1993), \\
P. Ginsparg in:
Les Houches (1988), Fields, Strings and Critical Phenomena.
\bibitem [17]{q} D. Li; Modern. Phys. Lett. {\bf B7} (1993) 1103.
\bibitem [18]{r} R. Dijkgraf, E. Verlinde, and H. Verlinde; Commun. Math. Phys
{\bf 115} (1988) 649.
\bibitem [19]{s} T. Eguchi and H. Ooguri; Nucl. Phys. {\bf B289} (1987) 308.
\bibitem [20]{t} H. T. Sato; Mod. Phys. Lett. {\bf A9} (1991) 451.
\bibitem [21]{u}  N. Aizawa, S. Sachse, and H. T. Sato; Mod. Phys. Lett.
{\bf A10} (1995) 853.
\bibitem [22]{v}  M. Alimohammadi and A. Shafei Deh Abad; J. Phys. A {\bf A29}
(1996) 559.
\bibitem [23]{w}  C. Gomez and G. Sierra; Lecture Notes in Physics
{\bf 375} (1990).
\bibitem [24]{x} J. E. Avron, M. Klein, and A. Pnueli; Phys. Rev. Lett. {\bf 69}
(1992) 128.

\end{thebibliography}
\end{document}